\documentclass[fleqn,usenatbib]{mnras}

\usepackage{newtxtext,newtxmath}

\usepackage[T1]{fontenc}
\usepackage{ae,aecompl}

\usepackage{graphicx}	
\usepackage{amsmath}	
\usepackage{amssymb}	
\usepackage{url}
\usepackage[normalem]{ulem}  

\title[Open cluster rotation in 3D]{Linking the rotation of a cluster to the spins of its stars: The kinematics of NGC~6791 and NGC~6819 in 3D}

\author[S.~Kamann et al.]{
S. Kamann,$^{1}$\thanks{E-mail: skamann@ljmu.ac.uk}
N.~J. Bastian,$^{1}$
M. Gieles,$^{2,3,4}$
E. Balbinot,$^{2,5}$
V. H\'{e}nault-Brunet$^{6}$
\\
$^{1}$Astrophysics Research Institute, Liverpool John Moores University, 146 Brownlow Hill, Liverpool L3 5RF, UK\\
$^{2}$Department of Physics, University of Surrey, Guildford, GU2 7XH, UK\\
$^3$Institut de Ci\`{e}ncies del Cosmos (ICCUB), Universitat de Barcelona, Mart\'{i} i Franqu\`{e}s 1, E08028 Barcelona, Spain\\
$^4$ICREA, Pg. Lluis Companys 23, 08010 Barcelona, Spain.\\
$^{5}$Kapteyn Astronomical Institute, University of Groningen, Landleven 12, 9747 AD Groningen, The Netherlands\\
$^{6}$National Research Council, Herzberg Astronomy \& Astrophysics, 5071 West Saanich Road, Victoria, BC, V9E 2E7, Canada
}

\date{Accepted XXX. Received YYY; in original form ZZZ}

\pubyear{2018}

\begin{document}
\label{firstpage}
\pagerange{\pageref{firstpage}--\pageref{lastpage}}
\maketitle

\begin{abstract}
The physics governing the formation of star clusters is still not entirely understood. One open question concerns the amount of angular momentum that newly formed clusters possess after emerging from their parent gas clouds. Recent results suggest an alignment of stellar spins and binary orbital spins in star clusters, which support a scenario in which clusters are born with net angular momentum cascading down to stellar scales. In this paper, we combine {\it Gaia} data and published line of sight velocities to explore if NGC~6791 and NGC~6819, two of the clusters for which an alignment of stellar spins has been reported, rotate in the same plane as their stars. We find evidence for rotation in NGC~6791 using both proper motions and line of sight velocities. Our estimate of the inclination angle is broadly consistent with the mean inclination that has been determined for its stars, but the uncertainties are still substantial. Our results identify NGC~6791 as a promising follow-up candidate to investigate the link between cluster and stellar rotation. We find no evidence for rotation in NGC~6819.
\end{abstract}

\begin{keywords}
stars: kinematics and dynamics -- open clusters and associations: individual: NG~6791, NGC~6819 -- astrometry -- techniques: radial velocities
\end{keywords}



\section{Introduction}
Recent results have suggested that the alignment of stellar rotation axes in clusters is not isotropically distributed, but rather has a preferential inclination angle \citep[][]{2017NatAs...1E..64C}.  This result is based on asteroseismology analysis of Kepler data for stars in NGC~6791 and NGC~6819, two Galactic open clusters with ages of $\sim8$~Gyr and $\sim2.4$~Gyr, respectively.  While some caution needs to be taken in interpreting these results, due to potential systematic effects in inferring the stellar inclination angle from Kepler analyses \citep{2018MNRAS.479..391K}, the results potentially offer an unexpected and intriguing insight into the formation of stellar clusters.  There has also been other studies that have found non-isotropic stellar orientations in clusters using photometric rotation periods and spectroscopic rotation velocities \citep[e.g.][]{2018A&A...612L...2K}.

In addition to individual stellar rotational orientations, alignment of binary orbital spins within clusters has been reported. \citet{2013ApJ...773...54K} studied the young ($\sim3$~Myr) cluster/H{\sc ii} region NGC~2264, and found that within the cluster the binary stars were preferentially seen edge-on (i.e. in eclipsing binaries) and that outside the cluster they were isotropically distributed.   

The generally advocated explanation for the alignment of stellar rotation axes (and aligned binary orbital spins) in clusters is that the stars were born within a giant molecular cloud (GMC) that contained significant amounts of angular momentum, which was inherited by the stars as they formed \citep[e.g.][]{2017NatAs...1E..64C,2018MNRAS.481L..16R}. If the alignment of stars/binaries is strong at birth it persists within a cluster, as extremely close, i.e. rare, stellar dynamical interactions are required in order to change the orientation of the spin of a star. For tidal friction to be efficient, an encounter must be within a few stellar radii \citep[e.g.][]{1977A&A....57..383Z}.

If either the stellar or the binary orientations are aligned, and if this is due to inheriting angular momentum from the progenitor GMCs, then we may expect that the host stellar cluster is itself rotating with the same orientation.  Coherent cluster rotation has been found in clusters of all ages, from recently formed \citep[e.g.][]{2012A&A...545L...1H} to intermediate age \citep[e.g.][]{2011MNRAS.411.1386D,2018MNRAS.480.1689K} and even in the ancient GCs \citep[e.g.][]{2012A&A...538A..18B,2014ApJ...787L..26F,2018MNRAS.473.5591K, 2018ApJ...860...50F}. \citet{2018MNRAS.473.5591K} further found a correlation between the angular momentum of the clusters and their relaxation times \citep[see also][]{2018MNRAS.481.2125B}. As relaxation causes a cluster to lose angular momentum via transport of angular momentum and escaping stars \citep[e.g.][]{1999MNRAS.302...81E}, this finding suggests that rotation is an intrinsic part of the cluster formation process. Such a scenario is also suggested by the detection of ordered motions in young \ion{H}{ii} regions \citep{2018MNRAS.478.3530D}.

The feasibility of studying the rotation of star clusters with the {\it Gaia} DR2 catalogue \citep{2016A&A...595A...1G,2018A&A...616A...1G} has already been demonstrated by \citet{2018MNRAS.481.2125B} and \citet{2018MNRAS.479.5005M}.  \citet{2018MNRAS.481.2125B} further inferred inclination angles for several clusters by comparing to results obtained from line of sight (LOS) velocities. Here, we take advantage of the {\it Gaia} data and published LOS velocity measurements to look for evidence of cluster rotation in NGC~6791 and NGC~6819, two of the clusters with reported stellar rotation alignment.  The goal is to see if the rotation axes of the clusters (if rotation is detected) align with the stellar rotational axes.

NGC~6791 \citep[$d=4.2\,{\rm kpc}$,][]{2011ApJ...729L..10B} is a peculiar cluster given its old age \citep[$\sim8\,{\rm Gyr}$, e.g.,][]{2008A&A...492..171G} and high metallicity \citep[${\rm [Fe/H]}=+0.4$, e.g.,][]{2006ApJ...643.1151C}. In light of these properties and its eccentric orbit, it has been suggested that it originated in the Galactic bulge \citep[e.g.][]{2006A&A...460L..27B,2012A&A...541A..64J}. \citet{2015MNRAS.449.1811D} observed tidal tails in NGC~6791 and concluded that the cluster must have initially been significantly more massive than its current mass of $5\,000\,{\rm M_\odot}$.

NGC~6819 \citep[$d=2.4\,{\rm kpc}$,][]{2011ApJ...729L..10B} is considerably younger \citep[$\sim2.4$~Gyr,][]{1998AJ....115.1516R} than NGC~6791 and has about half of its mass \citep[$2\,600\,{\rm M_\odot}$,][]{2001AJ....122..266K}. Its metallicity is about solar \citep[e.g.][]{2001AJ....121..327B} and its orbit \citep[][]{2013AJ....146...43P} consistent with an origin in the Galactic disk.

\section{Data}

\subsection{Line of sight velocities}
\label{sec:radial_velocities}
Both clusters, NGC~6791 and NGC~6819, have been targeted by the WIYN open cluster study \citep[WOCS,][]{2000ASPC..198..517M}. We made use of the published catalogues which include the average LOS velocities of the target stars, as well as their uncertainties and variability flags.

\citet{2014AJ....148...61T} provide data for 280 stars towards NGC~6791. While no uncertainties are provided for the velocities of the individual stars, they can be recovered from the average uncertainty of $0.38\,{\rm km\,s^{-1}}$ and the ratios between the standard deviation of each star and the overall precision \citep[$e/i$, cf. Table~1 in][]{2014AJ....148...61T}. After excluding stars flagged as non-members, possible members, and binary candidates, we were left with a sample of 101 stars.

For NGC~6819, we combined two WOCS studies, namely \citet{2009AJ....138..159H} and \citet{2015AJ....149..121L}, providing accurate velocities for 1\,207 and 333 stars, respectively. For 304 stars that we found to be present in both studies (based on a spatial offset $<10^{-3}\,{\rm arcmin}$) we obtained combined LOS velocities using weighted averages of the individual measurements. The standard deviation between the catalogues (after correcting for a global offset of $0.23\,{\rm km\,s^{-1}}$) was $1.06\,{\rm km\,s^{-1}}$, in line with the mean uncertainty of $0.79\,{\rm km\,s^{-1}}$ that we  obtained for the combined velocities. Exclusion of non-members, possible members, and binary stars resulted in a final sample of 397 LOS velocities for NGC~6819.

\subsection{{\it Gaia} photometry and astrometry}
\label{sec:gaia}

We selected all stars available in {\it Gaia} DR2 that lie within a radius of $20\arcmin$ from any of the two cluster centres. The size of the selection radius was motivated by the spatial extents of the clusters on the sky. For NGC~6791, \citet{2011ApJ...733L...1P} find a tidal radius of $23.1\arcmin$ using King profile fits and for NGC~6819, a Roche radius of $23\arcmin$ has been determined by \citet{2001AJ....122..266K}.

To select cluster members, each star was assigned a membership probability based on its parallax and proper motion measurements. In brief, for each of the three quantities ($x\in\{\varpi,\,\mu_{\rm \alpha}^*,\,\mu_{\rm \delta}\}$),  we defined a Gaussian probability density,
\[p(x) = \frac{1}{\sqrt{2\pi(\epsilon_{x}^2 + \sigma_{x}^2)}}\exp\left\{-\frac{(x-\bar{x})^2}{2(\epsilon_{x}^2 + \sigma_{x}^2)^2}\right\}\,,\]
where $\epsilon_{x}$ indicates the measurement uncertainty from {\it Gaia}, $\bar{x}$ is the mean value of the cluster, and $\sigma_{x}$ the intrinsic dispersion of the cluster. Cluster parallaxes were obtained by inverting the (uncorrected) distance moduli from \citet{2018ApJ...863...65C}, $(m-M) = 13.29$ (NGC~6791) and $12.22$ (NGC~6819).\footnote{Note that the distance moduli were determined by \citet{2018ApJ...863...65C} from the {\it Gaia} parallaxes of their cluster members, hence were not corrected for the systematic offset of $-0.029\,{\rm mas/yr}$ in DR2 \citep{2018A&A...616A...2L}.} We neglected the (small) intrinsic dispersions caused by the finite cluster widths, i.e. set $\sigma_{\rm \varpi}=0$. Further, we determined initial guesses for the systemic motions of the clusters visually by searching for overdensities of the full {\it Gaia} samples in proper motion space. The initial guesses were iteratively refined after a preliminary member selection had been performed and the final values are provided in Table~\ref{tab:cluster_properties}. They are in good agreement with previous estimates of the systemic proper motions of the clusters \citep[e.g.][]{2006A&A...460L..27B,2013AJ....146...43P}. As intrinsic dispersion, we assumed $1\,{\rm km\,s^{-1}}$, comparable to the virial estimates of both clusters. 

\begin{figure}
	\includegraphics[width=\columnwidth]{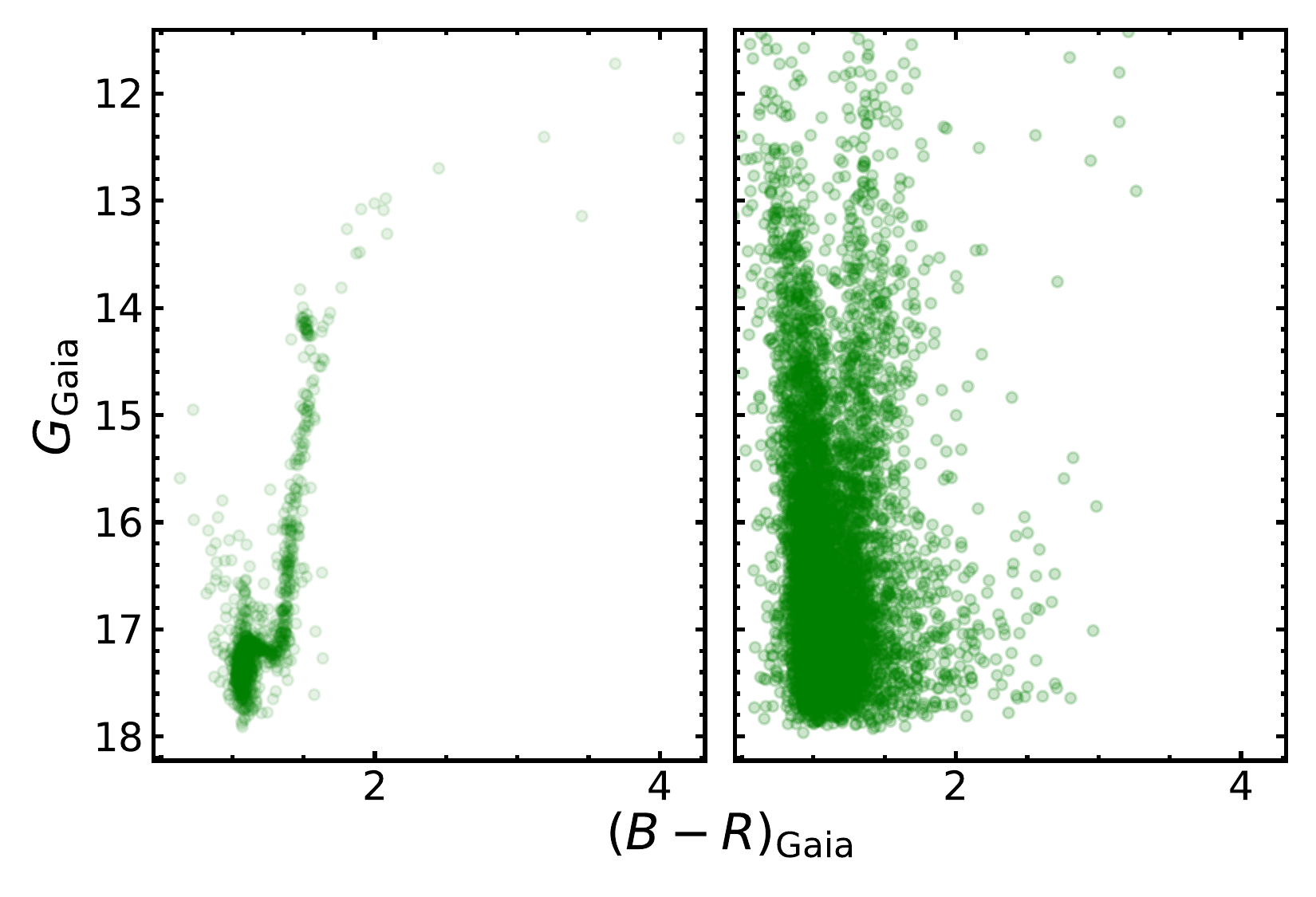}
    \caption{{\it Gaia} colour-magnitude diagrams for stars in direction of NGC~6791 (within a distance of $20\arcmin$ of the assumed cluster centre), separated into cluster members ({\it left}) and field stars ({\it right}).}
    \label{fig:ngc6791_gaia_cmd}
\end{figure}

\begin{figure}
	\includegraphics[width=\columnwidth]{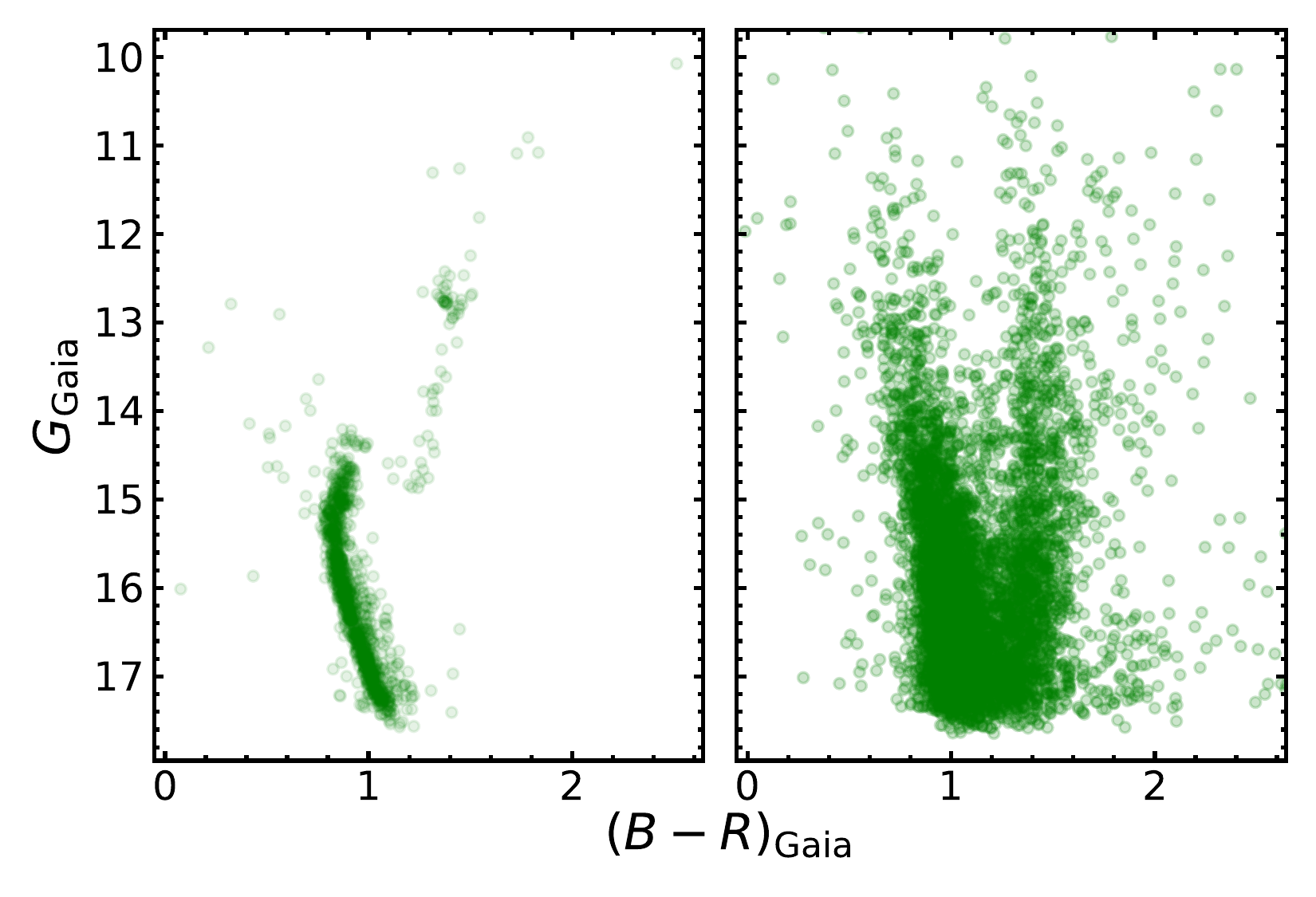}
    \caption{Same as Fig.~\ref{fig:ngc6791_gaia_cmd} for NGC~6819.}
    \label{fig:ngc6819_gaia_cmd}
\end{figure}

Each star with a combined membership probability of $p(\varpi)\times p(\mu_{\rm \alpha}^*)\times p(\mu_{\delta}) > 0.01$ was considered a cluster member. In addition, we only kept stars with precisely measured proper motions, corresponding to uncertainties $<4\,{\rm km\,s^{-1}}$ and $<2\,{\rm km\,s^{-1}}$ at the distances of NGC~6791 and NGC~6819, respectively. These error cuts were chosen as a compromise between obtaining sufficiently large samples of stars while avoiding contamination of our samples with stars having poorly determined kinematics. When using less stringent error cuts, we started to observe contamination by field stars in the colour-magnitude diagrams (cf. Figs~\ref{fig:ngc6791_gaia_cmd} and \ref{fig:ngc6819_gaia_cmd}). Note that the chosen error cuts correspond to approximately the same accuracy in proper motion space because the distance of NGC~6791 is about a factor of 2 larger than that of NGC~6819.

Our member selection resulted in samples of 964 and 1\,244 stars for NGC~6791 and NGC~6819, respectively. In order to illustrate the success of our membership selection, we show colour-magnitude diagrams of both clusters, split up into members and foreground stars, in Figs~\ref{fig:ngc6791_gaia_cmd} and \ref{fig:ngc6819_gaia_cmd}.

Finally, we note that DR2 contains only few LOS velocities for cluster members ($6$ in NGC~6791 and $35$ in NGC~6819). Therefore, we did not include {\it Gaia} LOS velocities in our analysis.

\section{Cluster morphologies}
\label{sec:morphologies}

The catalogs of cluster members created from the {\it Gaia} data allow us to determine the exact locations of the cluster centres and to study the two-dimensional density structures of the clusters. The latter is important because dynamical models usually require some assumptions on the morphology of the system (such as an axisymmetric cluster, see \citealt{2006A&A...445..513V}).

\subsection{Cluster centres}
In order to determine the locations of the cluster centres, we proceeded as follows. For each cluster, we defined a grid of $50\times50$ points of trial centres, with a total length of $2\,{\rm arcmin}$ on each side, around the centre positions available in SIMBAD. At each grid point, we selected the stars from our {\it Gaia} samples within a distance of $8\,{\rm arcmin}$ and split them up into eight equally sized pie slices, each covering an angle of $\pi/4$.  Then we compared the number of stars in the pie slices opposite to each other. Under the hypothesis that a grid point represents the true centre of a cluster, each star has equal probability of falling into either of the two slices. Hence, the number ratio of stars between the two slices would follow a binomial distribution and we can assign each grid point a probability of representing the cluster centre. The actual centre of each cluster was then obtained by fitting a two-dimensional Gaussian to the distribution of probabilities. The central coordinates we found this way are provided in Table~\ref{tab:cluster_properties}.

\begin{table}
	\centering
	\caption{Cluster properties derived in this work.}
	\label{tab:cluster_properties}
	\begin{tabular}{ccc} 
		\hline
		 & NGC~6791 & NGC~6819 \\
		\hline
		$\alpha$ (J2000) & 19:20:51.3 & 19:41:17.2 \\
		$\delta$ (J2000) & +37:46:26 & 40:11:18 \\
		$\mu_{\alpha}^*\,[{\rm mas/yr}]$ & -0.42 & -2.91 \\
        $\mu_{\delta}\,[{\rm mas/yr}]$ & -2.27 & -3.86 \\
        $r_{\rm hl}\,[{\rm arcmin}]$ & $4.10\pm0.16$ & $5.12\pm0.28$ \\
		\hline
	\end{tabular}
\end{table}

\subsection{Ellipticities and position angles}
\label{sec:ellipticities}

\begin{figure*}
 \includegraphics[width=\columnwidth]{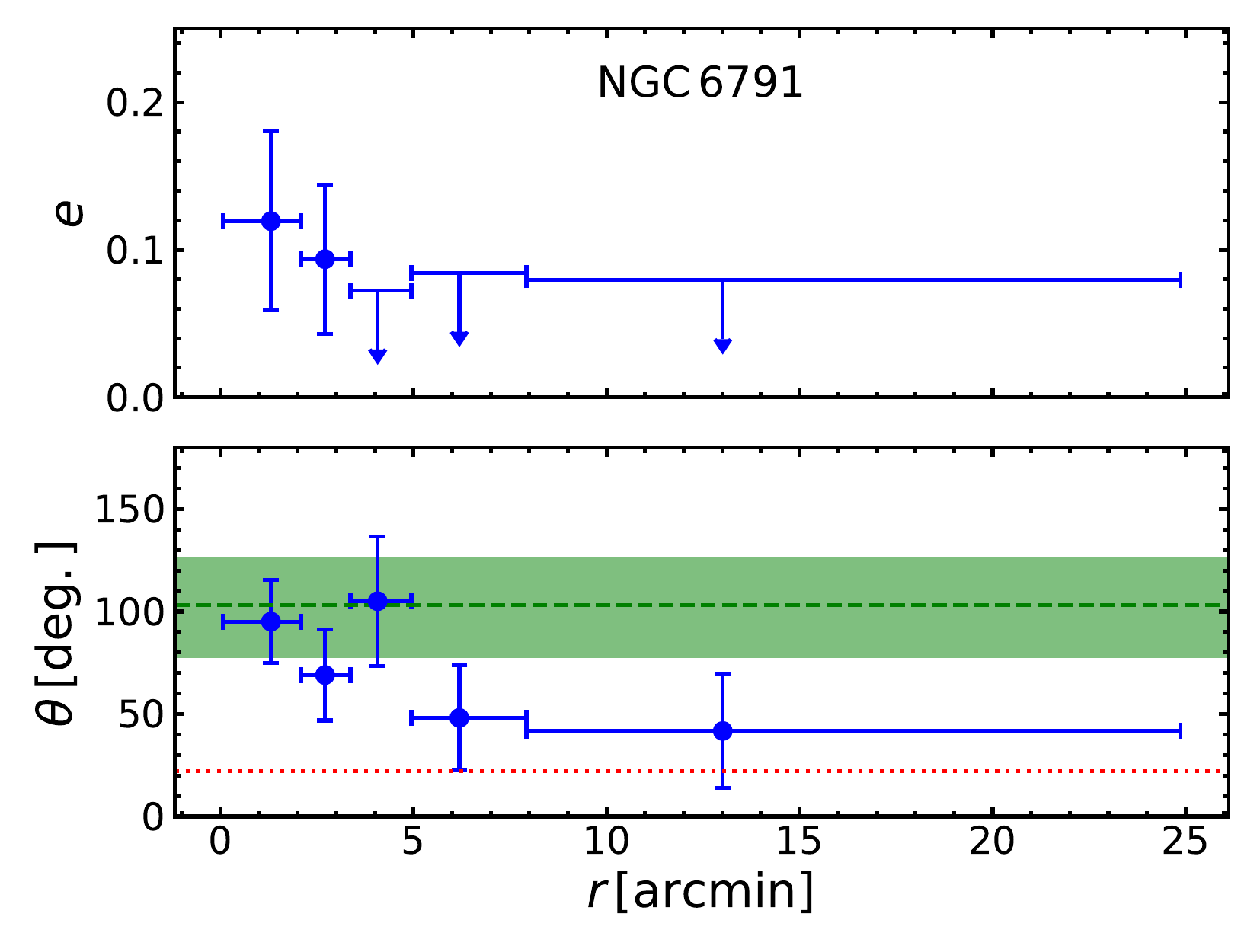}
 \includegraphics[width=\columnwidth]{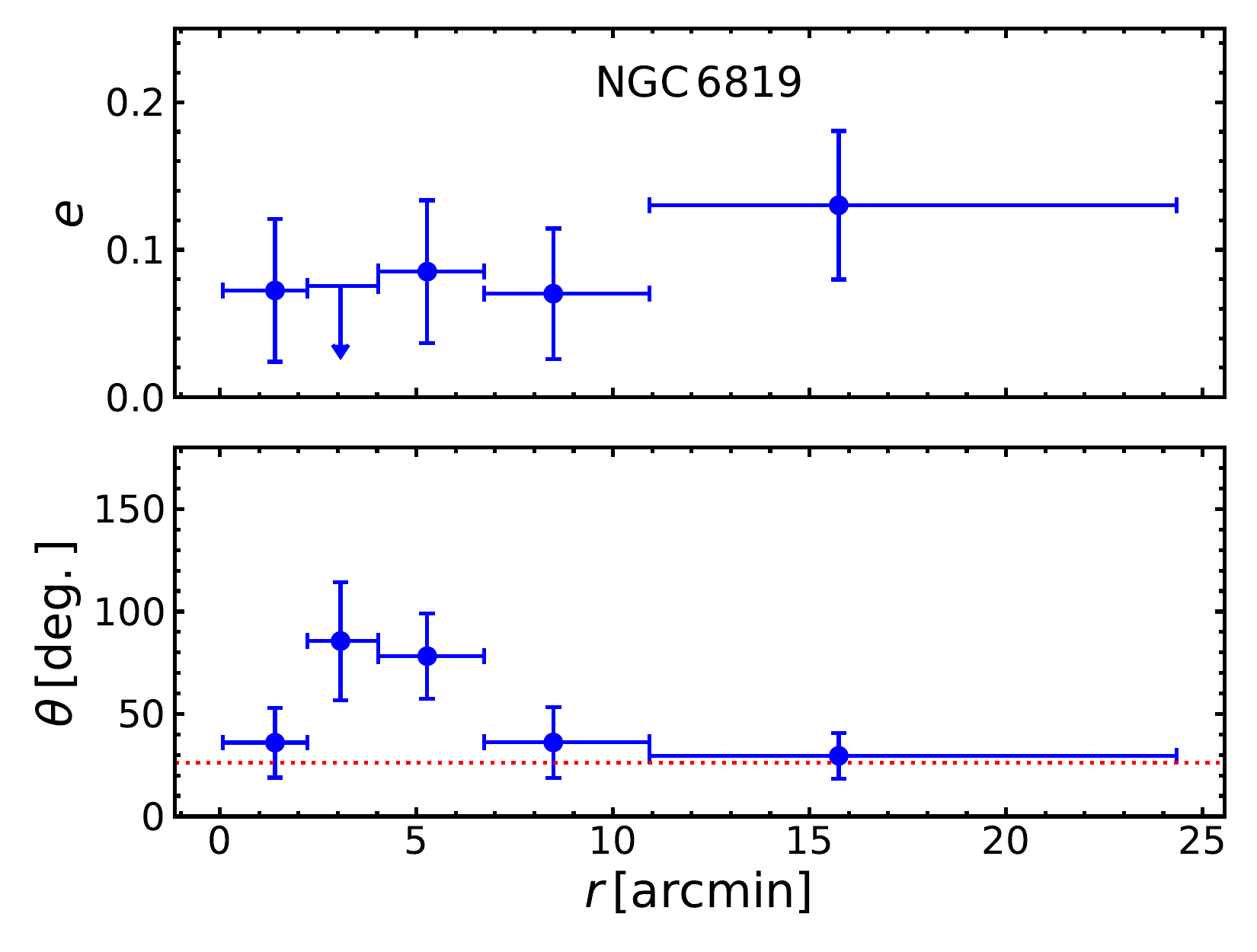}
 \caption{Ellipticity (\textit{top}) and position angle of the semi-major axis (\textit{bottom}, measured from north through east) of NGC~6791 (\textit{left}) and NGC~6819 (\textit{right}) as a function of distance to the cluster centre. The dotted red lines in the bottom panels indicate the directions towards the Galactic Centre. For NGC~6791, we also show the position angle expected from the analysis of the LOS velocities (green dashed line, cf. Sect.~\ref{sec:rv_analysis}). Horizontal error bars show the limits of the radial bins.}
 \label{fig:ellipticities}
\end{figure*}

With the centres of the clusters in place, we proceeded by constraining their ellipticities $e$ and position angles $\theta$. The latter provide the orientations of the semi-major axes and are measured from north through east, i.e. in counterclockwise direction. We followed \citet{2014ApJ...787L..26F} and first determined the eigenvalues $\lambda_{1,2}$ and eigenvectors $\vec{v}_{1,2}$ of the two-dimensional distributions of {\it Gaia} members of each cluster. The eigenvalues are related to the lengths of the semi-major and semi-minor axes via $\lambda_1 = 1/a^2$ and $\lambda_2=1/b^2$. Hence, the ellipticity $e$ of each cluster is given as $e=1-\sqrt{\lambda_1/\lambda_2}$. The eigenvector $\vec{v}_2$ points in the direction of the semi-major axis.

As the impact of external effects such as tidal forces will vary with distance to the cluster centres, we calculated the morphologies of NGC~6791 and NGC~6819 in different radial bins. The results of this calculation are shown in Fig.~\ref{fig:ellipticities}. The uncertainties included in Fig.~\ref{fig:ellipticities} were estimated by considering two effects that could potentially impact our measurements, namely uncertainties in our determination of the cluster centres and the finite number of stars per radial bin. We checked the influence of the former by varying the cluster centres within a radius of $5\,{\rm arcsec}$ (the approximate uncertainty of our centre determination) and recalculating the eigenvalues and eigenvectors. We found that our measurements were hardly affected by the exact locations of the cluster centres.

On the other hand, the finite number of stars seems to be the dominant source of uncertainty. To measure its influence, we created radial star count profiles and fitted them with multiple Gaussians \citep[using the code of][]{2002MNRAS.333..400C}. Then we created mock {\it Gaia} samples by randomly drawing stars from the multi-Gaussian model and assigning each star a random position angle from the interval $[0, 2\pi)$. For each of the mock samples, we determined the ellipticity and position angle. The standard deviations of the two distributions thus obtained served as estimates of the parameter uncertainties due to the finite number of stars. From the radial profiles, we further estimated half-light radii of the two clusters.  They are included in Table~\ref{tab:cluster_properties}.

We include in the lower panels of Fig.~\ref{fig:ellipticities} the directions towards the Galactic Centre, $\theta_{\rm GC}$, as dashed red lines.\footnote{Note that for both clusters, the Galactic Centre is actually located at $\theta_{\rm GC} - \pi$. However, our method to determine $e$ and $\theta$ is invariant to angular offsets of $\pi$.} In the bottom right panel of Fig.~\ref{fig:ellipticities}, it can be seen that for NGC~6819 there is a general agreement between $\theta_{\rm GC}$ and the position angle of the semi-major axis, indicating that the cluster is stretched along a line passing through the cluster and the Galactic Centre. The elongation further seems to increase with distance to the cluster centre, with the outermost bin having an ellipticity of $e\sim0.15$, whereas $e<0.1$ in the inner four bins, although the significance of this trend is low. Both of these effects, the stretch towards the Galactic Centre and its increase in the cluster outskirts, are expected for star clusters evolving in the Galactic tidal field \citep[][]{1987MNRAS.224..193T, 2001MNRAS.321..199P,2004AJ....128.2306C}. The strengths of these effects is expected to increase with cluster age. However, NGC~6791, which is significantly older than NGC~6819, does show a different behaviour. Its position angle is significantly offset from the direction towards the Galactic Centre and its ellipticity seems to \emph{decrease} with radius. This suggests that other effects, such as internal rotation or pressure anisotropies, impact the shape of NGC~6791 \citep[e.g.][]{2008AJ....135.1731V}.

At first sight, our results for NGC~6791 seem in tension with the findings of \citet{2015MNRAS.449.1811D}, who observed a clear elongation of NGC~6791 towards the Galactic Centre using ground based data. However, their Fig.~5 shows that this elongation is restricted to distances $>500\arcsec$ from the cluster centre, a radial regime in which our sample is lacking sufficient stars for a detailed analysis. At distances $<500\arcsec$, \citet{2015MNRAS.449.1811D} observe a shift of the position angle by $\sim90\,{\rm degrees}$, in good agreement with the behaviour visible in Fig.~\ref{fig:ellipticities}. We will revisit the peculiar elongation pattern of NGC~6791 in the context of cluster rotation below.

\section{Cluster dynamics}

Our analysis of the cluster dynamics is carried out in several steps. First, we analyse the LOS velocities and the proper motions of each cluster separately. If evidence for rotation is found, we then combine the results obtained from the two analyses to infer an inclination angle. In order to verify our approach, we perform the same analysis on a mock data set, extracted from an $N$-body model of a rotating cluster. The latter is described in detail in Appendix~\ref{app:mock_data}. Briefly, the mock data is based on the final snapshot of an $N$-body model of a rotating cluster evolved to an age of 10.75 Gyr, for which the initial dimensionless spin parameter $\lambda$ \citep{1969ApJ...155..393P} was set to $\lambda=0.091$ (model \textsc{mgen1} in Table 1 of \citealt{2015MNRAS.450.1164H}), corresponding to about 10\% of the total kinetic energy of the cluster in rotation at the start of the simulation.

\subsection{Line of sight velocities}
\label{sec:rv_analysis}

\begin{figure*}
	\includegraphics[width=\columnwidth]{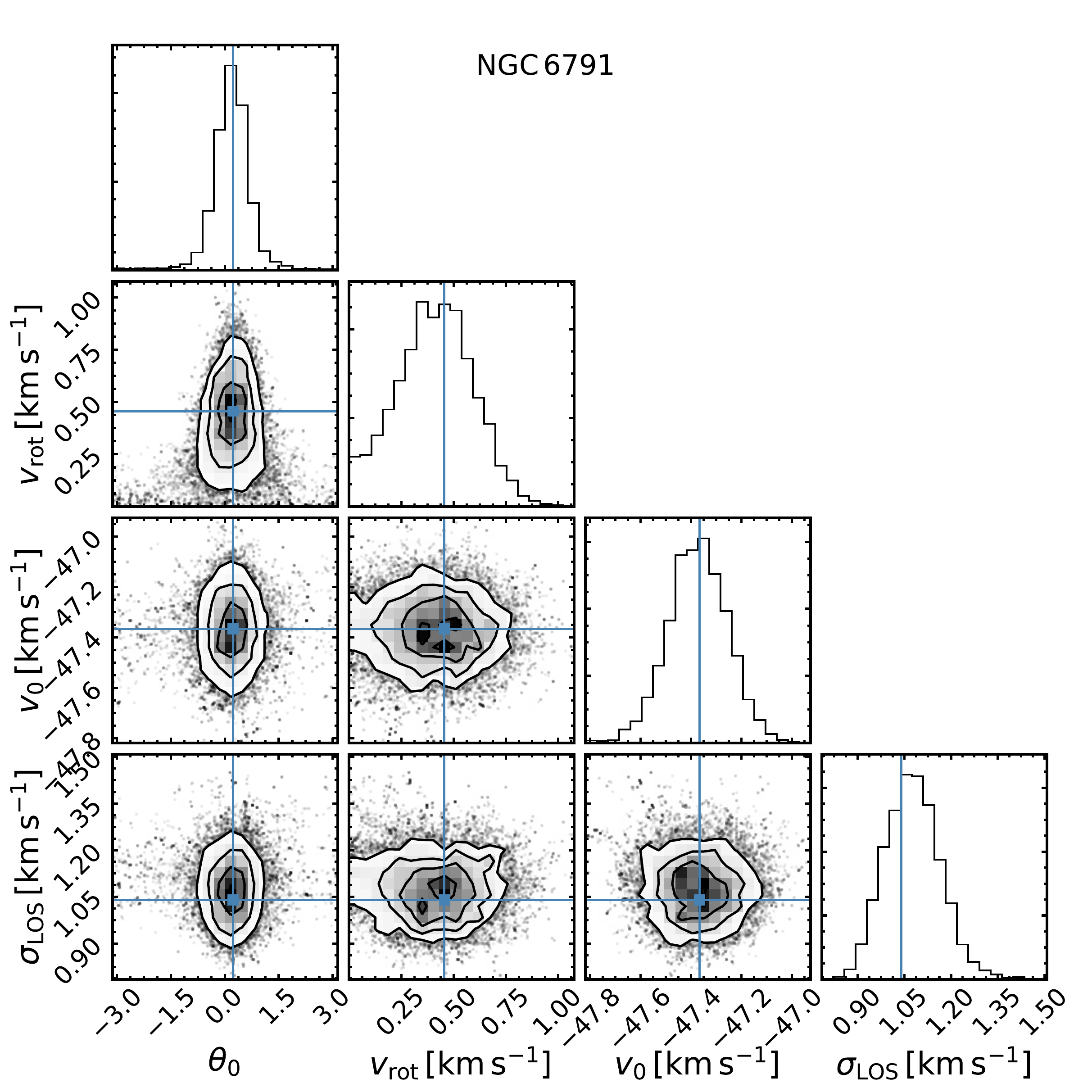}
    \includegraphics[width=\columnwidth]{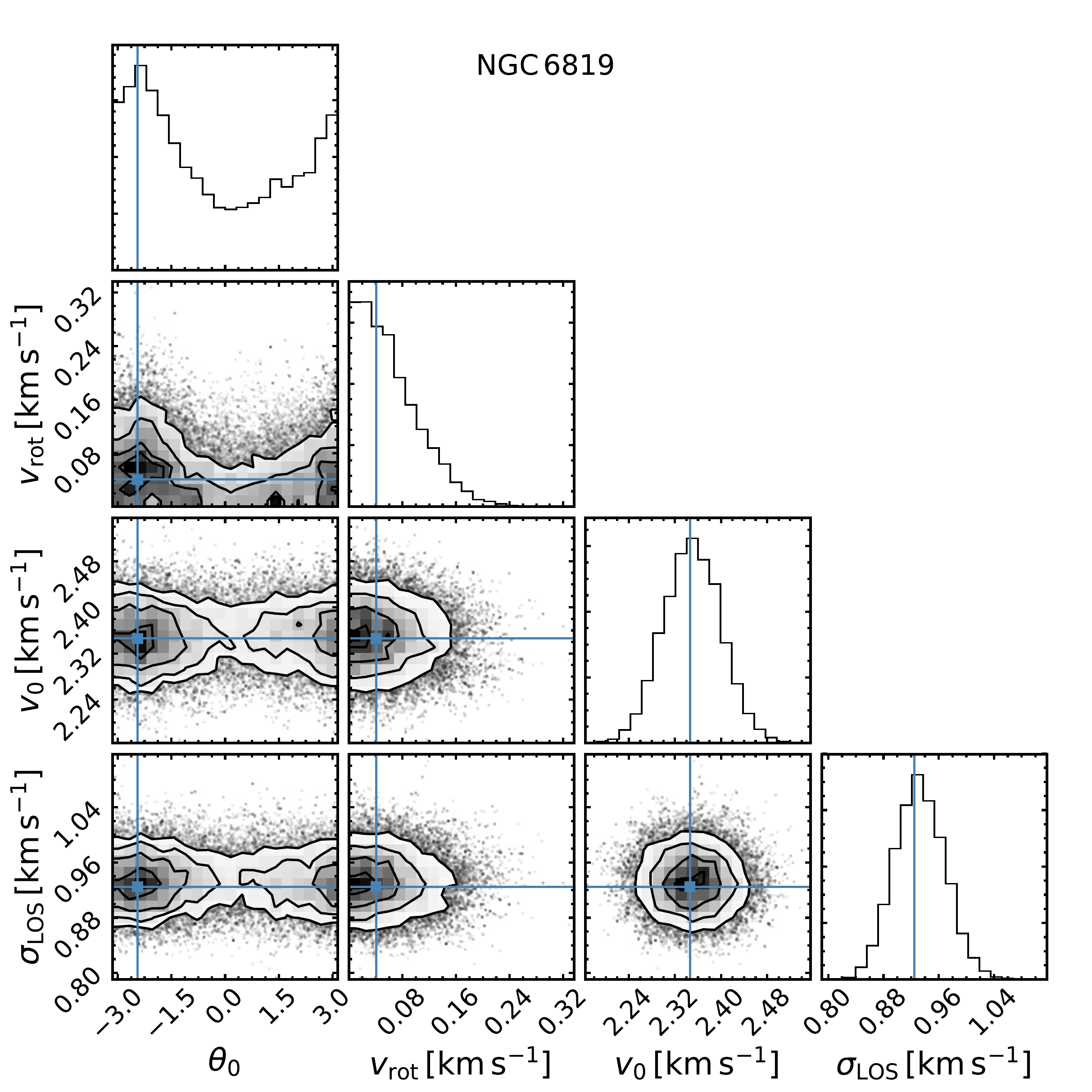}
    \caption{Kinematic cluster parameters of NGC~6791 (\textit{left}) and NGC~6819 (\textit{right}) obtained from an MCMC analysis of the LOS velocities as described in the text. The top panel of each column shows the 1D histogram of parameter values sampled by the MCMC walkers, whereas subjacent panels show 2D distributions of parameter pairs. The median value of each parameter is indicated by a solid blue line.}
    \label{fig:cornerplot}
\end{figure*}

We followed the approach outlined in \citet{2018MNRAS.473.5591K} to obtain maximum-likelihood estimates of the systemic velocity $v_{\rm 0}$, the velocity dispersion $\sigma_{\rm LOS}$, the rotation velocity $v_{\rm rot}$, and the position angle $\theta_{\rm 0}$ of the rotation axis for each cluster at the same time. In brief, it works by maximizing the likelihood $\mathcal{L}$, calculated according to,
\begin{equation}
\ln \mathcal{L} = -\sum_{i=1}^{n}\ln\left(2\pi\sqrt{\sigma_{\rm LOS}^2 + \epsilon_i^2}\right) - \frac{(v_i - v_{\rm c})^2}{2\,(\sigma_{\rm LOS}^2 + \epsilon_i^2)}\,,
\label{eq:likelihood}
\end{equation}
where $v_{\rm c}=v_{\rm 0}+v_{\rm rot}\,\sin(\theta_i-\theta_{\rm 0})$ is the systemic cluster velocity corrected for the effect of rotation. For each of the $n$ sources entering the above equation, the position angle $\theta_i$, the LOS velocity $v_i$, and its uncertainty $\epsilon_i$ must be known. Note that {equation~(\ref{eq:likelihood}) assumes constant dispersion and rotation curves. However, the generalization to spatially varying curves can be easily accomplished using analytical functions that evaluate $\sigma_{\rm LOS}(R_i,\theta_i)$ and $v_{\rm rot}(R_i,\theta_i)$ for each stellar position, where $(R_i, \theta_i)$ are the positions of the stars on the plane of the sky. Yet in view of the relatively small sample sizes, especially for NGC~6791, we did not try to obtain radial profiles for the various parameters but rather concentrated on obtaining global values. A caveat with this approach is that the global values could be biased towards the dispersion and rotational amplitude in the region of the cluster best sampled by the data. However, the half-number radii of our LOS velocity samples are $4.1\arcmin$ (NGC~6791) and $4.8\arcmin$ (NGC~6819), respectively, very similar to the half-light radii of the clusters (cf. Table~\ref{tab:cluster_properties}). Therefore, we do not expect that assuming a constant rotation curve is dragging our results in a certain direction because of preferential coverage.

Prior to the analysis, we corrected the LOS velocities for the effect of perspective rotation, using the systemic proper motions of the clusters listed in Table~\ref{tab:cluster_properties} and the method outlined in \citet{2006A&A...445..513V}. The corrections we applied to the individual velocities reached from $-0.15\,{\rm km\,s^{-1}}$ to $0.27\,{\rm km\,s^{-1}}$ for NGC~6791 and from $-0.20\,{\rm km\,s^{-1}}$ to $0.19\,{\rm km\,s^{-1}}$ for NGC~6819.

In order to assign uncertainties to each parameter, we used the code \textsc{ emcee} of \citet{2013PASP..125..306F} and obtained the 16th and 84th percentiles of the parameter posterior probability distributions resulting from 100 MCMC walkers with 500 steps each. For both clusters, best-fit values (defined as the median values of the posterior distributions) and the uncertainties of the four model parameters are provided in Table~\ref{tab:radial_velocities}. Further, Fig.~\ref{fig:cornerplot} shows corner plots of the parameter distributions returned by the MCMC sampler. The plots were created using the code of \citet{corner}.

In NGC~6791, we detect rotation at the $2.2\sigma$ level, around an axis offset by 14.2 degrees from north in counterclockwise direction. The rotation velocity of $0.40\,{\rm km\,s^{-1}}$ corresponds to about 37 per cent of the measured velocity dispersion. This $v_{\rm rot}/\sigma_{\rm LOS}$ ratio is similar to the one we find in the analysis of the mock data used to validate our fitting method (cf. Appendix~\ref{app:mock_data}, see Fig.~\ref{fig:mock_cornerplot}).

To investigate if the rotation of NGC~6791 leaves an imprint in the morphology of the cluster, we overplot in the lower left panel of Fig.~\ref{fig:ellipticities} the position of the semi-major axis one would expect if the elongation of the cluster was caused by rotation\footnote{Note that for an oblate rotator the semi-major axis is perpendicular to the rotation axis. This seems to be the case for most globular clusters \citep{2014ApJ...787L..26F,2018MNRAS.473.5591K}.}. The elongation of the inner three bins agrees very well with this expectation whereas the orientation of the outer bins is intermediate between those expected from rotation and tidal forces. So the inner parts of NGC~6791 appear to behave as an oblate rotator while tides may also influence the shape of the cluster beyond $5\arcmin$.

On the other hand, we do not find any signs of rotation along the line of sight in NGC~6819, despite the fact that our sample of stars is almost 4 times as large compared to NGC~6791. Instead, we obtain a very stringent upper limit of $0.11\,{\rm km\,s^{-1}}$ for the rotation amplitude, cf. Table~\ref{tab:radial_velocities}. Given the similarity in $\sigma_{\rm LOS}$ for the two clusters, this implies that either NGC~6819 has a lower intrinsic rotational velocity, or its rotation is more in the plane of the sky. We will address this in the next section.

\subsection{{\it Gaia} proper motions}
\label{sec:pm_analysis}

To study the cluster dynamics using the {\it Gaia} data, we defined a Cartesian coordinate system for each cluster, with its origin on the cluster centre, $x$ increasing westwards, and $y$ increasing northwards. The world coordinates and proper motions from the {\it Gaia} catalog were transformed into the new coordinate system using standard formulas \citep[see, e.g.,][]{2018arXiv180409381G}. Afterwards, we converted the relative proper motions to polar coordinates. The formulas for this conversion can be found in, e.g., \citet{2000A&A...360..472V}. The uncertainties of $\mu_{\rm \alpha}^*$ and $\mu_{\delta}$ were propagated during this process while possible covariances between the five astrometric parameters were neglected.

\begin{table}
	\centering
	\caption{Dynamical cluster parameters derived from the line of sight velocities.}
	\label{tab:radial_velocities}
	\begin{tabular}{ccc} 
		\hline
		 & NGC~6791 & NGC~6819 \\
		\hline
		$v_{\rm 0}\,[{\rm km\,s^{-1}}]$ & $-47.37^{+0.11}_{-0.12}$ & $2.35^{0.05}_{0.05}$ \\
		$\sigma_{\rm LOS}\,[{}\rm km\,s^{-1}]$ & $1.07^{+0.10}_{-0.08}$ & $0.93^{0.04}_{0.04}$ \\
		$v_{\rm rot}\,[{\rm km\,s^{-1}}]$ & $0.40^{+0.17}_{-0.18}$ & $0.05^{0.06}_{0.04}$ \\
        $\theta_{\rm 0}\,[{\rm deg.}]$ & $14^{+24}_{-26}$ & $-153^{+91}_{-98}$ \\
		\hline
	\end{tabular}
\end{table}

Polar coordinates offer a convenient way to study rotation using proper motions because the intrinsic rotation is mainly contained in the tangential component while the perspective expansion or contraction effect caused by the LOS velocity of the system only affects the radial component. Hence, in order to obtain the cluster kinematics, we applied a similar maximum-likelihood approach as for the LOS velocities (cf. Sect.~\ref{sec:rv_analysis}) independently to the tangential and the radial proper motion component. However, this time we only solved for the mean value $\langle \mu \rangle$ and the dispersion $\sigma_\mu$ of each component, i.e. set $v_{\rm c}=v_{\rm 0}$ in equation~(\ref{eq:likelihood}), as even for a rotating cluster no sinusoidal velocity variation with position angle is expected. Instead, the mean value of the tangential component corresponds to the rotation velocity in the plane of the sky. Thanks to the samples of $\sim1\,000$ stars with reliable proper motion measurements available in {\it Gaia} DR2, we were able to create radial profiles of the proper motion kinematics. They are displayed in Fig.~\ref{fig:pm_profiles}.

\begin{figure*}
	\includegraphics[width=\columnwidth]{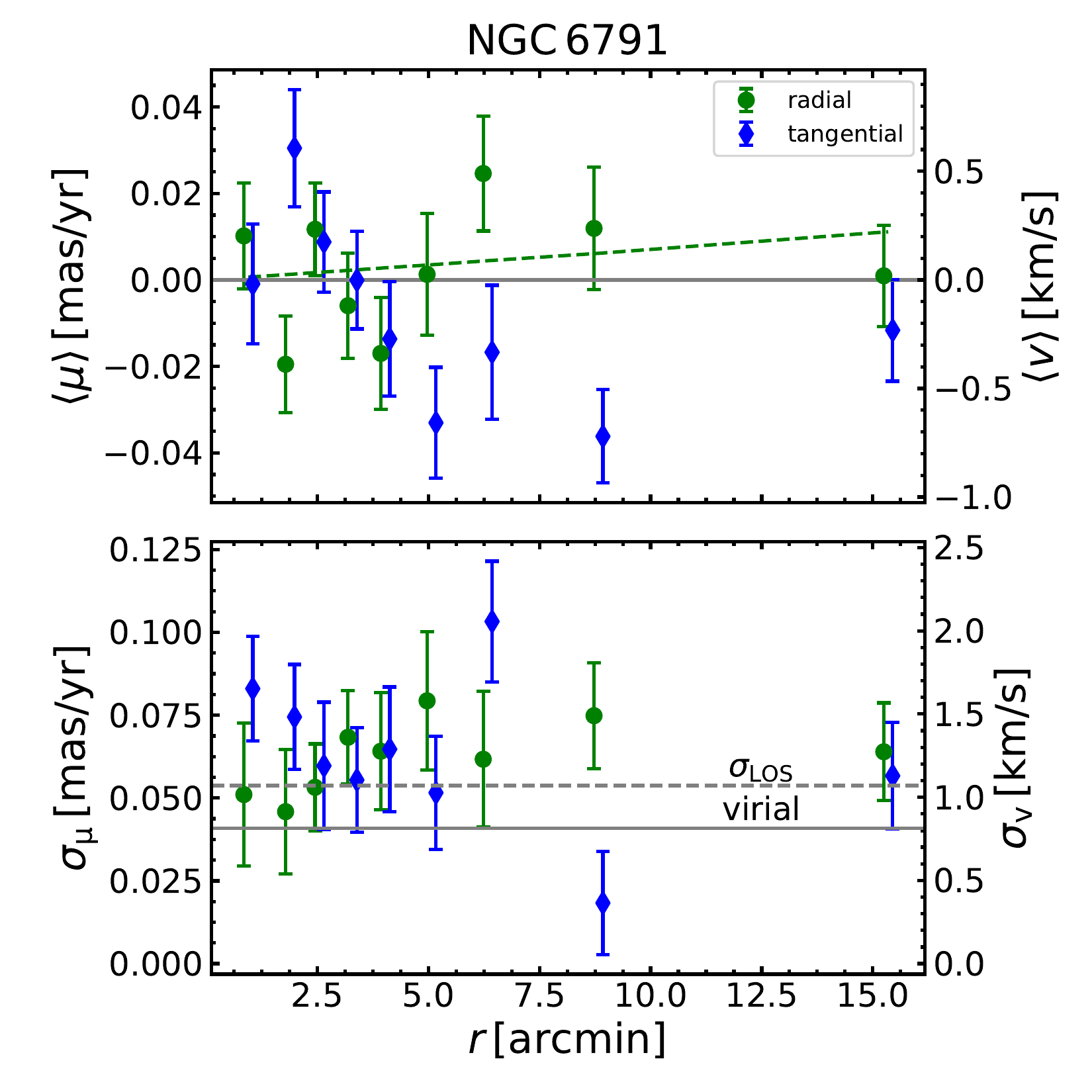}
    \includegraphics[width=\columnwidth]{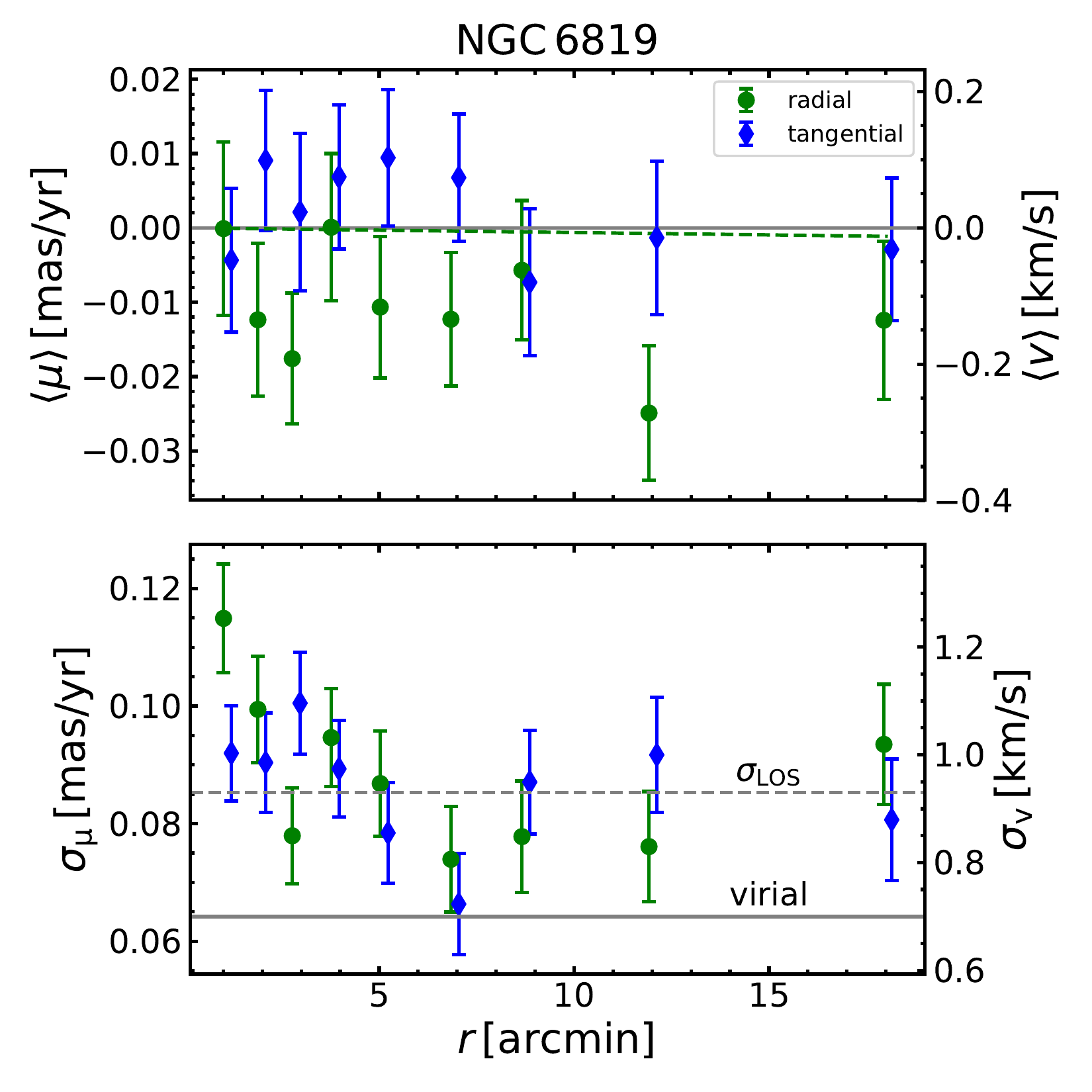}
    \caption{Proper motion kinematics of NGC~6791 (\textit{left}) and NGC~6819 (\textit{right}). The radial profiles of the mean velocities are shown in the {\it top} panels while the {\it bottom} panels show the velocity dispersion profiles as a function of distance to the cluster centres. Blue diamonds show the results obtained for the tangential components while green circles are used for the results obtained for the radial components. In the upper panels, a dashed green line indicates the expected behaviour of the radial component according to the LOS velocity of the cluster (perspective contraction or expansion). The grey solid line in the lower panels indicates the virial estimate of the velocity dispersion of each cluster while the grey dashed line indicates the dispersion value obtained from the LOS velocities. The distance moduli of NGC~6791 and NGC~6819 provided by \citet{2011ApJ...729L..10B} have been used to convert the measured proper motions from ${\rm mas\,yr^{-1}}$ to ${\rm km\,s^{-1}}$ as indicated on the right y-axis of each panel.}
    \label{fig:pm_profiles}
\end{figure*}

In the upper panels of Fig.~\ref{fig:pm_profiles}, dashed green lines indicate the expected expansion or contraction of each cluster given its systemic velocity. The predictions were obtained by expressing the first two lines of equation~6 in \citet{2006A&A...445..513V} in polar coordinates. The mean values of the radial component in NGC~6791 are in reasonable agreement with the prediction. A mild expansion is observed, as expected given its approaching LOS velocity of $-47.4\,{\rm km\,s^{-1}}$. For NGC~6819, a very small contraction is expected from its receding LOS velocity of $2.4\,{\rm km\,s^{-1}}$. However, the contraction that we observe is stronger. An almost constant offset of $\sim0.01\pm0.01\,{\rm mas\,yr^{-1}}$ ($\sim0.1\pm0.1\,{\rm km\,s^{-1}}$) is visible in the upper right panel of Fig.~\ref{fig:pm_profiles}. The reason for this behaviour is currently unknown, but it seems unlikely that the cluster is actually contracting. More likely, we are looking at systematic effects in the {\it Gaia} data. We will revisit this issue in Sect.~\ref{sec:systematics} below.

The radial profiles of the mean tangential proper motions depicted in the upper panels of Fig.~\ref{fig:pm_profiles} confirm the results found using the LOS velocities. While no significant deviations from zero are observed for NGC~6819, the mean values in NGC~6791 show a trend towards negative values at distances $>4\arcmin$ to the cluster centre. This trend corresponds to clockwise rotation.

The mean tangential proper motion outside of $4\arcmin$ is $-0.022\pm0.006\,{\rm mas\,yr^{-1}}$. At an assumed distance of $4.2\,{\rm kpc}$, this corresponds to a rotation velocity of $0.42\pm0.12\,{\rm km\,s^{-1}}$ on the sky, i.e. comparable to the rotational amplitude seen in the line-of-sight velocities. However, this comparison neglects that we assumed a constant rotation velocity when analysing the LOS velocities. If we do the same for the proper motions, we get a mean tangential velocity of $-0.007\pm0.004\,{\rm mas\,yr^{-1}}$, corresponding to $0.14\pm0.08\,{\rm km\,s^{-1}}$.

Interestingly, the radial trend observed for the proper motions seems to be absent in the LOS velocities. If we split the latter sample in two subsamples according to the distance to the cluster centre (lower and higher than $4\,{\rm arcmin}$), both subsamples yield a rotation velocity of around $0.4\,{\rm km\,s^{-1}}$. However, the sample sizes are small, so that a potential radial trend may also be hidden in the noise. Finally, we note that no rotation signal is detected in the proper motions of our LOS velocity stars (83/101 stars are in the {\it Gaia} sample). The $1\sigma$ upper limit of $0.008\,{\rm mas\,yr^{-1}}$ that we find corresponds to $0.16\,{\rm km\,s^{-1}}$.

We show the velocity dispersion profiles in the lower panels of Fig.~\ref{fig:pm_profiles}. Also shown in those panels are the velocity dispersions obtained during the analysis of the LOS velocity samples (cf. Sec.~\ref{sec:rv_analysis}) and predicted using simple virial estimates \citep[eq.~4.249b in][]{2008gady.book.....B},
\begin{equation}
 \langle v^2 \rangle = \frac{0.45\,G\,M_{\rm c}}{r_{\rm hl}},
\end{equation}
with $\langle v^2 \rangle=3\sigma^2$. We assumed cluster masses of $M_{\rm c}=5\,000\,{\rm M_\odot}$ \citep[NGC~6791,][]{2015MNRAS.449.1811D} and 
$2\,600\,{\rm M_\odot}$ \citep[NGC~6819,][]{2001AJ....122..266K}. The half-light radii $r_{\rm hl}$, listed in Table~\ref{tab:cluster_properties}, were derived from the star count profiles discussed in Sect.~\ref{sec:ellipticities}. The virial estimate agrees reasonably well with the measured dispersion in NGC~6791 and underestimates our measurements in NGC~6819. Such an offset could be caused by underestimated velocity uncertainties. The median uncertainties of our proper motion samples correspond to $2.4\,{\rm km\,s^{-1}}$ (NGC~6791) and $0.8\,{\rm km\,s^{-1}}$ (NGC~6819) and are comparable to the intrinsic dispersions, so they can potentially impact our dispersion measurements. However, for both clusters we find good agreement with the dispersion values determined from the LOS velocity data, indicating that the uncertainties for the {\it Gaia} proper motions are fairly accurate.

Nevertheless, deviations due to inaccurate uncertainties are still likely. An indication in this regard may be the lack of a decrease of the dispersion measurements with distance to the cluster centres, as one would expect for a relaxed cluster. Alternatively, a significant contribution from energetically unbound stars \citep[so-called potential escapers,][]{2010MNRAS.407.2241K,2017MNRAS.466.3937C,2017MNRAS.468.1453D} could explain this observation. In fact, the $N$-body models of NGC~6791 by \citet{2018MNRAS.474...32M} show that the velocity dispersion profile in projection decreases only by about 30 per cent in the central $\sim7\arcmin$ and stays almost constant beyond this radius. Note that we observe a similar behaviour in our analysis of the mock data presented in Appendix~\ref{app:mock_data}, (see Fig.~\ref{fig:mock_pm_profiles}).

Given the small internal velocities and potential residual errors in the {\it Gaia}  data, it is not possible to use the ratio between the tangential and radial dispersion profiles to make robust statements about possible anisotropies in the clusters. We only note that the ratios inferred from Fig.~\ref{fig:pm_profiles} are essentially all consistent with one, suggesting isotropy, as is expected for a cluster in the late stages of its evolution \citep[e.g.][]{2016MNRAS.462..696Z}.

\subsection{Inclination angles}
\label{sec:inclination}

Given the absence of any significant rotation signal in NGC~6819, we focus on NGC~6791 in our efforts to constrain the inclination angle of the rotation field. As a first estimate, we simply determined the arcus tangent of the ratio of the rotation velocities along the line of sight ($0.40\pm0.18\,{\rm km\,s^{-1}}$, cf. Table~\ref{tab:radial_velocities}) and in the plane of the sky ($-0.14\pm0.08\,{\rm km\,s^{-1}}$, cf. Sec.~\ref{sec:pm_analysis}). In doing so, we essentially treat the cluster as a rotating disk, seen at an inclination angle $i$. The comparison yields an inclination angle of $i=73^{+11}_{-28}\,{\rm deg}$ and hence suggests a strongly inclined rotation field.\footnote{Note that $i$ can be positive or negative, corresponding to counterclockwise or clockwise rotation in the plane of the sky. However, in the work by \citet{2017NatAs...1E..64C} only positive spin inclinations are considered, hence we also restrict ourselves to $i>0$.} To obtain a conservative lower limit for the inclination, we recalculate it using the mean tangential proper motion measured at distances $>4\arcmin$ to the cluster centre, $0.42\pm0.12\,{\rm km\,s^{-1}}$. This yields $i=43\pm15\,{\rm deg.}$. Finally, we also checked what happens if we restrict our proper motion sample to those stars that also have LOS velocity measurements. As already mentioned in Sec.~\ref{sec:pm_analysis}, no rotation is obvious in their proper motion distribution. Accordingly, we obtain a lower limit on the inclination angle of $i>68\,{\rm degrees}$.

A more sophisticated way to determine the inclination of the rotation field of a cluster was used by \citet{2006A&A...445..513V}. They used a relation originally found by \citet{1994MNRAS.271..202E} between local means of the proper motion along the semi-minor axis $\langle \mu_{\rm v}\rangle$ and the LOS velocity $\langle v_{\rm r}\rangle$ of an axisymmetric system,
\begin{equation}
\langle v_{\rm r}\rangle = 4.74\,d\,\tan i\langle \mu_{\rm v}\rangle\,,
\label{eq:pm_rv}
\end{equation}
where $d$ is the distance in kpc. In Sections~\ref{sec:morphologies} and ~\ref{sec:rv_analysis}, we found that the observed position angle of the photometric semi-major axis of NGC~6791 agrees reasonably well with the expectation for an oblate rotator. Hence, we assume in the following that the semi-major axis of NGC~6791 is perpendicular to the rotation axis, for which the position angle is provided in Table~\ref{tab:radial_velocities}. Under this assumption, we can rotate our Cartesian coordinate system into a new system ($u$, $v$) such that $u$ is aligned with the semi-major axis of the cluster. Following \citet{2006A&A...445..513V}, we then defined a polar grid in the new coordinate system and binned the data according to their coordinates in the grid. Before binning the data, we exploited the symmetries of the system to transform all velocity measurements to the first quadrant of the ($u$, $v$) coordinate system.\footnote{In an axisymmetric system, LOS velocities and minor-axis proper motions are symmetric to reflections about the semi-major axis and antisymmetric to reflections about the semi-minor axis. Major-axis proper motions are antisymmetric to reflections about the semi-major axis and symmetric to reflections about the semi-minor axis.} Given the relatively small number of LOS velocity measurements in NGC~6791, we selected the bin sizes such that each bin contained on average 10 such measurements.

Figure~\ref{fig:ngc6791_inclination_fit} shows the relation between the mean minor-axis proper motion and the mean LOS velocity for each bin in the polar grid. Based on our previous analyses, the mean LOS velocities are expected to be $>0$, while the mean minor-axis proper motions are expected to be $<0$ (corresponding to clockwise rotation). While the former can be verified from Fig.~\ref{fig:ngc6791_inclination_fit}, we only see mild evidence for the latter. However, a more obvious trend  might be hidden under the noise given that the means in all bins are consistent with negative values within their uncertainties.

We used an orthogonal distance regression method to fit the relation defined in eq.~\ref{eq:pm_rv} to the data shown in Fig.~\ref{fig:ngc6791_inclination_fit}. The method takes into account the uncertainties in both coordinates. We forced the fit to go through the centre of the coordinate system. It yielded an inclination angle of $45.4_{-18.0}^{+11.1}$ degrees.

For the stellar spins in NGC~6791, \citet{2017NatAs...1E..64C} found a mean inclination angle of $30\,{\rm deg.}$, with the most inclined stars having values of $\sim45\,{\rm deg.}$. Thus all of our estimates suggest that the rotation of the cluster is more strongly inclined than the mean stellar spin. However, some of our estimates are still consistent with $i=30\,{\rm deg.}$ within their uncertainties. In particular, this is the case for the value of $i$ we obtained using eq.~\ref{eq:pm_rv} and our analysis of the mock data (cf. Appendix~\ref{app:mock_data}) suggests that this value is the most robust. Therefore we cannot exclude the possibility that the cluster does rotate in the same plane that has been reported for the stars by \citet{2017NatAs...1E..64C}.

The uncertainties in our inclination estimates are still substantial. One reason for this is the relatively low number of LOS velocities, so that each bin in Fig.~\ref{fig:ngc6791_inclination_fit} contains only a few stars with such measurements. Accordingly, the uncertainties of the mean values are still comparable to the expected rotation signal. While the mean proper motions suffer much less from low number statistics, the uncertainties of the mean values derived from them are only a factor $\lesssim2$ lower. The reason is that the uncertainties of each measurement are larger than for the LOS velocities. This will likely improve with future {\it Gaia} data releases, as the end-of-mission uncertainties for {\it Gaia} proper motions are expected to be lower by a factor of $\geq5$ compared to DR2.

\begin{figure}
	\includegraphics[width=\columnwidth]{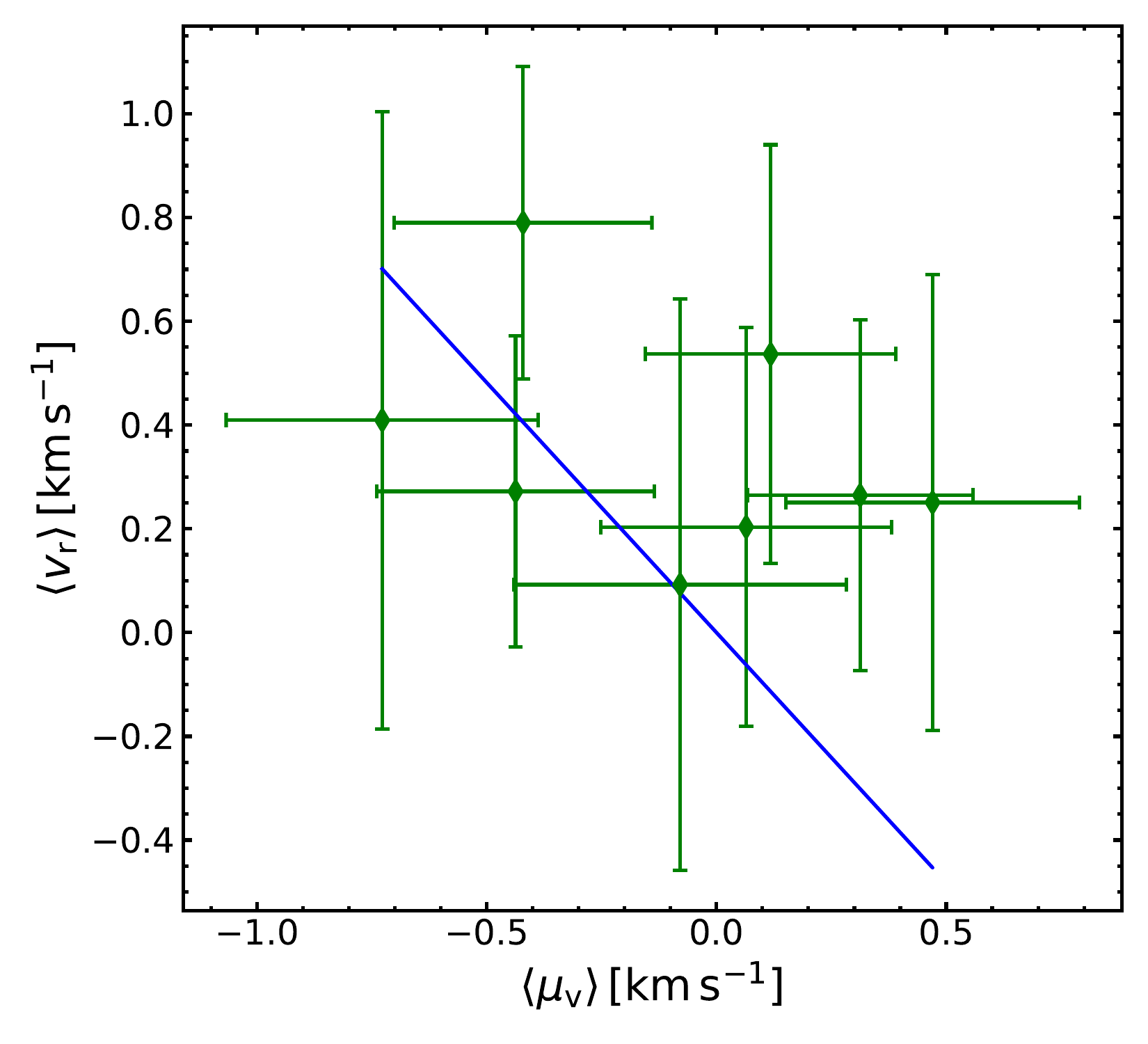}
    \caption{Relation between the mean proper motion along the assumed semi-minor axis and the mean LOS velocity for NGC~6791. The mean values have been obtained by binning the data in on a polar grid. The blue line shows a linear fit to the data, constrained to pass through the origin.}
    \label{fig:ngc6791_inclination_fit}
\end{figure}

\subsection{Systematic effects}
\label{sec:systematics}

The {\it Gaia} astrometric data suffers from well documented systematics \citep[e.g.][]{2018A&A...616A...2L,2018arXiv180409381G,2018MNRAS.481.2125B}. In order to verify if they have the potential to impact our results, we follow an approach outlined on the {\it Gaia} DR2 website\footnote{\url{https://www.cosmos.esa.int/web/gaia/dr2-known-issues}, see link to presentation by Lindegren et~al.} to determine the variance $\sigma_{\rm h}^2$ of the mean proper motion of a star cluster. According to this approach, the variance has a random and a systematic component and can be calculated according to the following formula for a sample of $n$ stars that have proper motion uncertainties $\sigma_{{\rm \mu}, i}$.
\begin{equation}
 \sigma_{\rm h}^2 = \frac{1}{n}\left(\frac{k}{n}\sum_{i=1}^{n} \sigma_{{\rm \mu}, i}^2 + V_{\rm \mu}(0)\right) + \frac{2}{n^2}\sum_{i=2}^{n}\sum_{j=1}^{i-1}V_{\rm \mu}(\phi_{i,j})\,.
 \label{eq:systematics}
\end{equation}
$k$ is a correction factor for the uncertainties provided in DR2 that we set to $1.0$. The spatial covariance function $V_{\rm \mu}$ has been derived by \citet{2018A&A...616A...2L} and can be retrieved from the {\it Gaia} webpage. It depends on the angular distance $\phi_{i,j}$ between two sources, given in degrees. The second, systematic term in equation~(\ref{eq:systematics}) is expected to dominate the error budget for large sample sizes.

Using the positions and proper motion uncertainties of our sample stars in both clusters, we found that the systematic errors affecting the proper motion means of the clusters are $\sim0.05\,{\rm mas\,yr^{-1}}$. A comparison with the upper right panel of Fig.~\ref{fig:pm_profiles} shows that this potential systematic error is larger than the apparent contraction visible in the dataset of NGC~6819. This indicates that with the current {\it Gaia} data it is not possible to verify if the observed trend is intrinsic or spurious. 

The potential systematic errors are also comparable to the rotation signal that we detect in NGC~6791 (cf. upper left panel in Fig.~\ref{fig:pm_profiles}). However, in this case it is reassuring that a rotation signal of similar strength is observed in the LOS velocity sample. Therefore, we do not expect that the rotation signal in NGC~6791 is caused by systematics in the data, although a final judgment will only be possible with future {\it Gaia} data releases.

\section{Conclusions}

In this work, we combined {\it Gaia} data with archival LOS velocities to study the internal dynamics of two old open clusters, NGC~6791 and NGC~6819, in three dimensions. For both clusters, an alignment of stellar spins has recently been reported \citep{2017NatAs...1E..64C}, suggesting that they inherited significant angular momentum from their parent gas clouds during formation. Such a transfer of angular momentum should also leave a fingerprint in the internal kinematics of the clusters and the aim of our study was to search for such fingerprints.

We found the clusters to have remarkably different kinematics. While both the LOS velocity and the proper motion data reveal clear signs of rotation in NGC~6791, no such signatures are found in NGC~6819. The rotation velocities we measure in NGC~6791 are $\sim0.5\,{\rm km\,s^{-1}}$ whereas our upper limits in NGC~6819 are $\lesssim0.1\,{\rm km\,s^{-1}}$. Both clusters have velocity dispersions $\sim1\,{\rm km\,s^{-1}}$, hence rotation seems far more important in NGC~6791 than in NGC~6819.

Star clusters experience significant mass loss due to tidal forces or their dynamical evolution. Stars escaping the clusters carry angular momentum, so that primordial rotation fields are expected to decrease with time \citep[e.g.][]{1999MNRAS.302...81E}.  We used eq.~(7.108) in \citet{2008gady.book.....B} to estimate relaxation times of 560~Myr (NGC~6791) and 260~Myr (NGC~6819), respectively. Hence, both clusters have cycled through about 10 relaxation times during their lives. Therefore, it is perhaps more surprising that NGC~6791 rotates than that NGC~6819 does not. One possibility is that NGC~6819 formed more compact and has therefore undergone more relaxation than the estimate from above.

We note that NGC~6791 is more massive than NGC~6819 ($5\,000\,{\rm M_\odot}$ compared to $2\,600\,{\rm M_\odot}$). Furthermore, \citet{2015MNRAS.449.1811D} suggest that NGC~6791 was born with a much higher mass of $(1.5-4)\times10^5\,{\rm M_\odot}$, although note that the initial mass of the $N$-body model of NGC~6791 by \citet{2018MNRAS.474...32M} was much lower ($5\times10^4\,{\rm M}_\odot$). A possibility to explain the differences in the dynamics of the two clusters may therefore be that rotation plays a more important role in high-mass clusters. Such a mass dependency could explain why rotation is observed in massive clusters of all ages \citep[e.g.][]{2012A&A...545L...1H,2011MNRAS.411.1386D,2018MNRAS.473.5591K} while \citet{2018arXiv180702115K} did not observe rotation in the young low-mass clusters of their sample.

The observed rotation velocities in NGC~6791 suggest a specific angular momentum comparable to that of rotating globular clusters with similar ages but much longer relaxation times \citep{2018MNRAS.473.5591K,2018MNRAS.481.2125B}. A possible explanation -- especially considering the mass loss that NGC~6791 likely experienced -- could be that part of the rotation signal is not primordial but has been introduced by tidal forces \citep[e.g.][]{2017MNRAS.466.3937C,2018ApJ...865...11L}. \citet{2016MNRAS.461..402T} found that in such cases a cluster would be characterized by solid-body rotation (i.e. constant angular velocities), which seems in line with the behaviour we observe in NGC~6791 (cf. Fig.~\ref{fig:pm_profiles}).

Given the properties of NGC~6791 (mass, size, orbit), tidally induced rotation is expected to become important at radii $\gtrsim10\arcmin$. In both datasets, LOS velocities and proper motions, we observe that rotation sets in further inwards. Nevertheless, we note that in the proper motion sample, we only find a significant rotation signal beyond $\sim4\arcmin$. But tides give rise to a net retrograde rotation (in a frame corotating with the cluster orbit) with respect to the orbit \citep[see][and references therein]{2017MNRAS.466.3937C}. Because the line-of-sight to NGC~6791 is approximately parallel to its orbital plane, this implies that they impact the LOS velocities stronger than the proper motions.

We determine an inclination angle of $i=45_{-18}^{+11}$ degrees for the cluster rotation field in NGC~6791. Taken at face value, it suggests that the cluster rotates in a plane that is titled with respect to the mean stellar spin orientation found by \citet{2017NatAs...1E..64C}. However, the uncertainties are still considerable, both for the stellar spin measurements \citep[e.g.][]{2018MNRAS.479..391K} and for our constraint on $i$. In addition, if the rotation field of the cluster is shaped by tidal forces, its orientation will likely be different from any primordial rotation field. Nevertheless, NGC~6791 occurs as a promising target for follow-up studies on a possible link between cluster rotation and stellar spins using future {\it Gaia} data releases and potentially larger LOS velocity samples. In addition, we plan to extend our studies to younger clusters where the effects of relaxation only play a minor role. In light of the findings by \citet{2018A&A...612L...2K}, M44 might be a promising targets as it is significantly younger, but also less massive, than NGC~6791 and NGC~6819.

\section*{Acknowledgements}
We thank the anonymous referee for a timely and well justified report that helped us to improve the paper.
We thank Joel Pfeffer for enlightening discussions during the preparation of the manuscript.
SK and NB gratefully acknowledge funding from a European Research Council consolidator grant (ERC-CoG-646928-Multi-Pop). MG acknowledges financial support from the Royal Society (University Research Fellowship) and MG and EB thank the European Research Council (ERC StG-335936, CLUSTERS) for financial support. EB acknowledges partial support by a Vici grant from the Netherlands Organisation for Scientific Research (NWO). 
VHB acknowledges support from the NRC-Canada Plaskett Fellowship. We also thank the International Space Science Institute (ISSI, Bern, CH) for welcoming the activities of the Team 407 ``Globular Clusters in the {\it Gaia} Era" (team leaders VHB \& MG), during which part of this work was conducted.
This research has made use of the SIMBAD database, operated at CDS, Strasbourg, France \citep{2000A&AS..143....9W}.
This work has made use of data from the European Space Agency (ESA) mission
{\it Gaia} (\url{https://www.cosmos.esa.int/gaia}), processed by the {\it Gaia}
Data Processing and Analysis Consortium (DPAC,
\url{https://www.cosmos.esa.int/web/gaia/dpac/consortium}). Funding for the DPAC
has been provided by national institutions, in particular the institutions
participating in the {\it Gaia} Multilateral Agreement.




\bibliographystyle{mnras}
\bibliography{bib}



\appendix

\section{Analysis of mock data}
\label{app:mock_data}

To verify if our approach can be used to get a useful constraint on the inclination angle in NGC~6791, we used a simulated cluster from the study of \citet{2015MNRAS.450.1164H}. The data is based on the final snapshot of an $N$-body model of a cluster evolved to an age of 10.75 Gyr, for which the initial dimensionless spin parameter $\lambda$ \citep{1969ApJ...155..393P} was set to $\lambda=0.091$ (model \textsc{mgen1} in Table 1 of \citealt{2015MNRAS.450.1164H}), corresponding to about 10\% of the total kinetic energy of the cluster in rotation at the start of the simulation. 

We scaled the simulation to the properties of NGC~6791 and extracted LOS velocity and proper motion samples from the simulation. This was done under the assumption that the rotation axis of the cluster is observed at a rotation axis angle of $\theta_{\rm 0}=-153\,{\rm deg.}$ (measured north through east) and at an inclination of $i=53\,{\rm deg.}$. The samples were chosen such that they had the same numbers of stars and the same radial distributions as the observed data of NGC~6791. Each velocity drawn from the simulation was then assigned the uncertainty of an actually observed star and scattered accordingly. Then, we performed the same analysis on the mock data as we did for the actual data of NGC~6791.

The results obtained for the LOS velocity data are summarized in Fig.~\ref{fig:mock_cornerplot}. It is obvious that the posterior probability distributions obtained by MCMC sampling show clear evidence for a rotating cluster. We obtain a rotation axis angle of $\theta_{\rm 0}=-146\pm19\,{\rm deg.}$, which is consistent with the rotation axis angle used to create the mock sample. We further obtain an amplitude of $v_{\rm rot}=0.51\pm0.17\,{\rm km\,s^{-1}}$ for the rotation velocity.

\begin{figure}
	\includegraphics[width=\columnwidth]{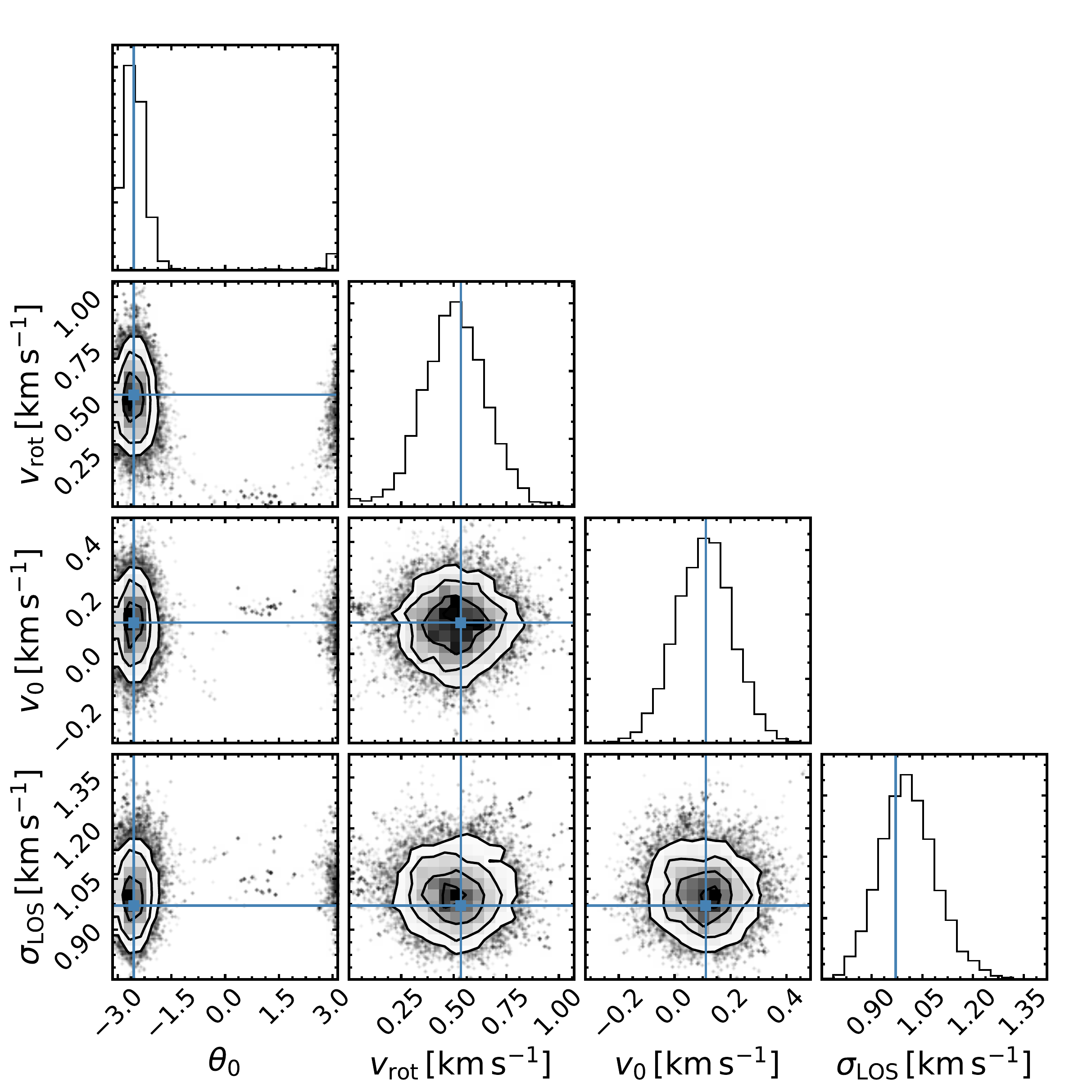}
    \caption{The same as Fig.~\ref{fig:cornerplot} for the mock data resembling NGC~6791.}
    \label{fig:mock_cornerplot}
\end{figure}

In the next step, we repeated the analysis performed for the proper motions. Fig.~\ref{fig:mock_pm_profiles} shows radial profiles of the mean velocity and the velocity dispersion for the tangential and the radial components of the proper motion. The tangential component shows clear signs of counterclockwise rotation, indicated by positive mean values. For the whole sample, we obtain a mean of $\mu_{\rm t}=0.66\pm0.10\,{\rm km\,s^{-1}}$ using the assumed distance of $4.92\,{\rm kpc}$. A direct comparison to the rotation amplitude obtained from the line of sight velocity data yields an estimate of the inclination angle of $i=38\pm10\,{\rm deg}$. which is $2\sigma$ off from the true value of $53$~degrees.

Next, we checked whether we can use eq.~\ref{eq:pm_rv} to obtain a more physically motivated estimate of the inclination angle. To this aim, we transformed our data to a coordinate system ($u$, $v$), where the $v$-axis is aligned with the semi-minor axis (or rotation axis) of the system found in the analysis of the LOS velocities. Afterwards, we binned the data on a polar grid (after exploiting the symmetry relations from Sect.~\ref{sec:inclination} to transform all measurements to the first quadrant) such that each grid point contained at least 10 LOS velocity measurements. Then, we obtained mean values for the LOS velocity and both proper motion components in each bin. The relation between the mean LOS velocity ($\langle v_{\rm r}\rangle$) and the mean of the minor-axis proper motion ($\langle \mu_{\rm v}\rangle$) for the individual bins is shown in Fig.~\ref{fig:mock_inclination}. Again, the blue line shows a linear fit to the data, performed with an orthogonal distance regression method that accounts for the uncertainties along both axes. Using equation~(\ref{eq:pm_rv}), we obtain an inclination angle of $i=51^{+7}_{-10}\,{\rm degress}$, in good agreement with our input value of $i=54$~degrees. Hence we are confident that the inclination angle determined this way for the actual data of NGC~6791 is robust.

\begin{figure}
	\includegraphics[width=\columnwidth]{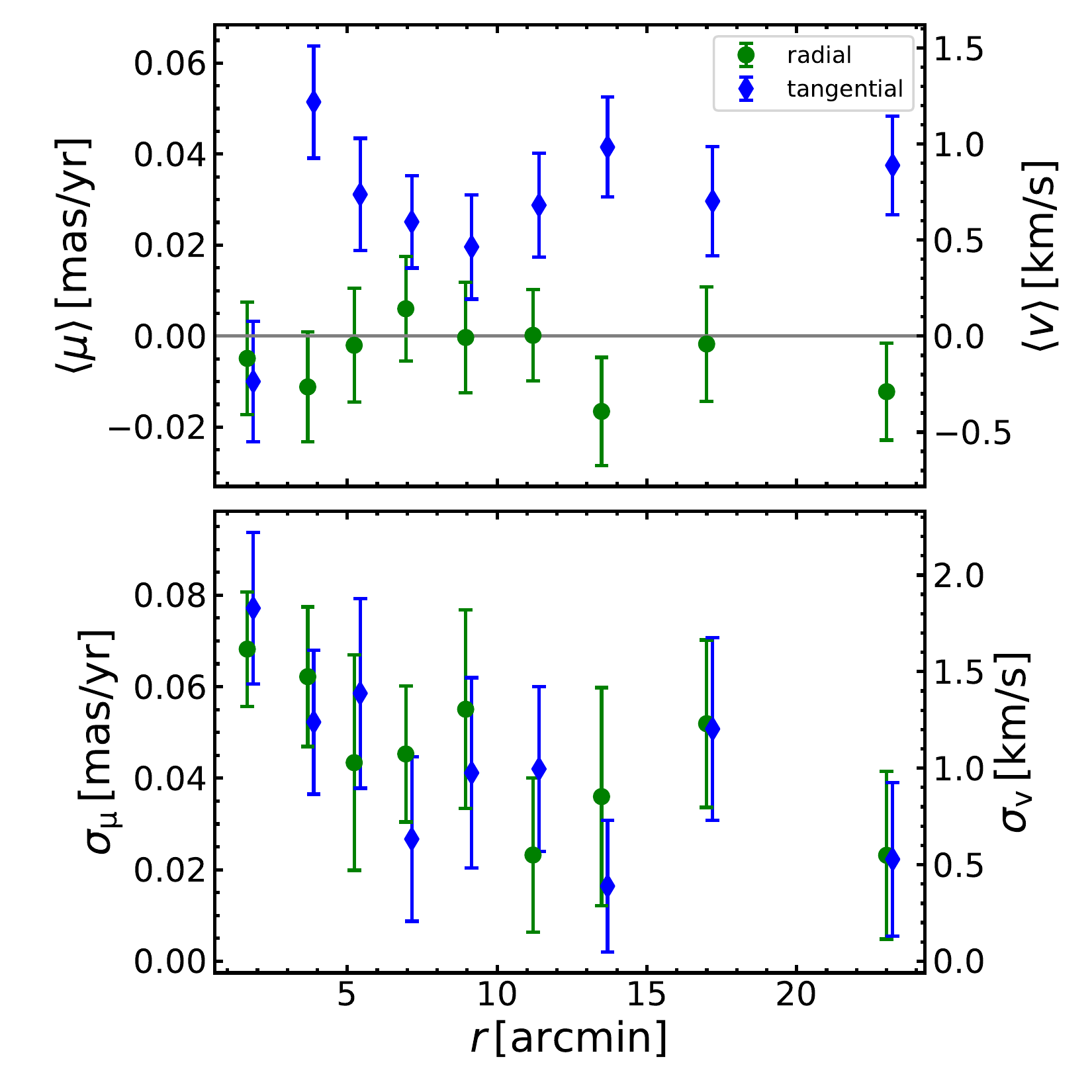}
    \caption{The same as Fig.~\ref{fig:pm_profiles} for the mock data resembling NGC~6791.}
    \label{fig:mock_pm_profiles}
\end{figure}

\begin{figure}
	\includegraphics[width=\columnwidth]{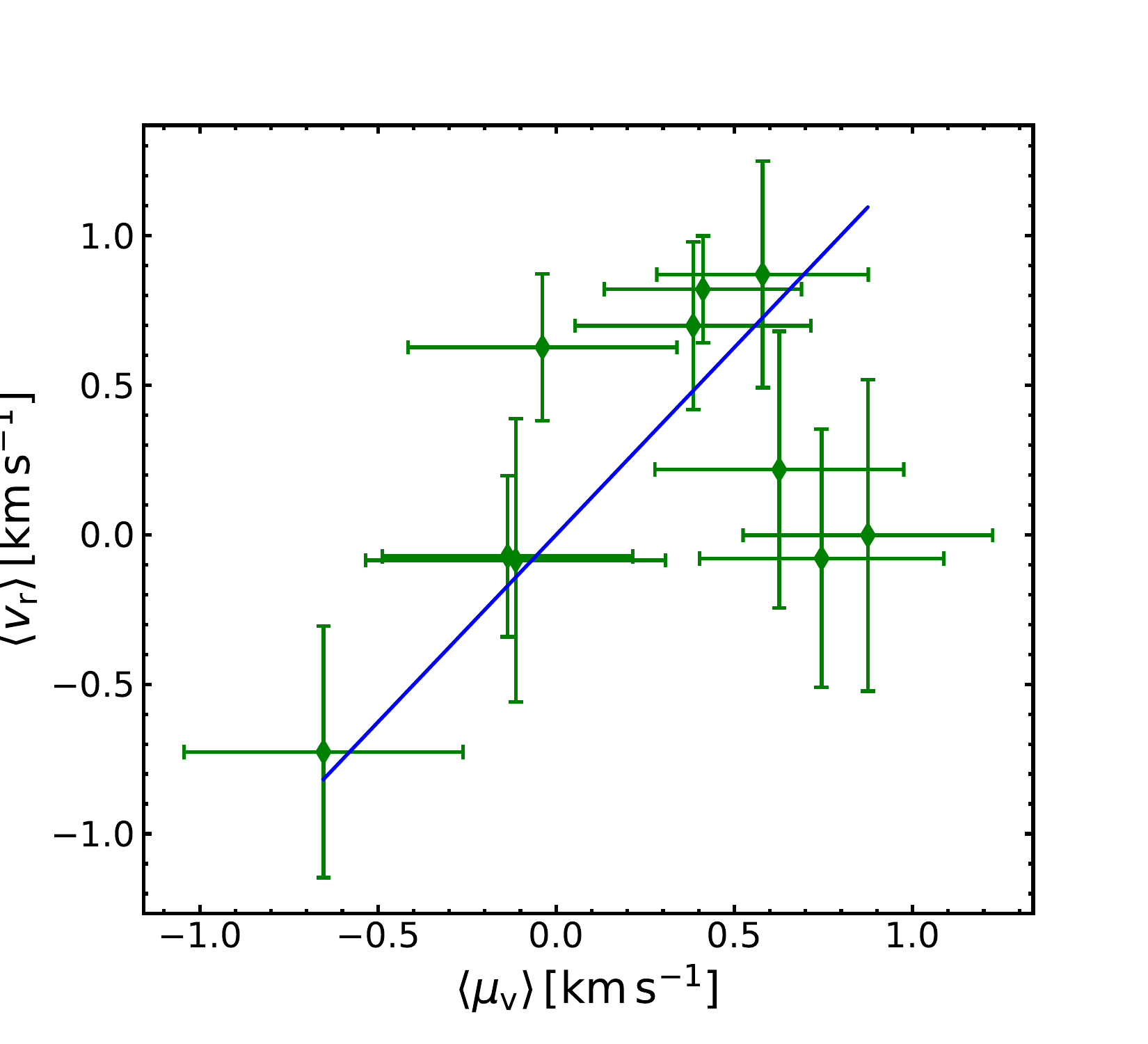}
    \caption{As in Fig.~\ref{fig:ngc6791_inclination_fit}, we show the relation between the mean of the LOS velocity and the mean of the proper motion component along the semi-minor axis in different spatial bins of the mock system. The blue line shows a linear fit to the data using an orthogonal distance regression method that takes into account the uncertainties in both coordinates. It was forced to go through the origin.}
    \label{fig:mock_inclination}
\end{figure}


\bsp	
\label{lastpage}
\end{document}